\documentstyle[twocolumn,prl,aps,floats,epsf]{revtex}

\begin{document}
\draft
\title{Strong-coupling approach to ground-state properties 
       of the Anderson lattice-model}
\author{Jan Brinckmann\cite{sces96-address}}
\address{Institut f{\"u}r Festk{\"o}rperphysik, Technische Hochschule
         Darmstadt, Germany}

\date{ 15 August 1996; to appear in Physica B }

\maketitle
\begin{abstract}
%
A Slave-Boson perturbational approach to ground-state properties of
the $U\to\infty$ periodic Anderson model is derived as an expansion
around the Atomic Limit ($V=0$)\,. In the case of zero temperature any
constraint-integral or 
limiting procedure can be avoided, a gauge-symmetry broken Mean-Field
phase is not involved. Physical quantities like the wave 
vector $k$ dependent Green's function obey a direct representation
in Feynman-skeleton diagrams in $k$-space. A
self-consistent $1/N$-expansion is derived, and its relation to the
limit of infinite spatial dimension $d\to\infty$ is pointed out.
\noindent

%
%
\end{abstract}
%
%
\newlength{\mysize} 
\def\loadepsfig#1{
 \def\figname{#1}
 \vbox to 10pt {\ }
 \vbox{ \hbox to \hsize {
   \mysize\hsize    \advance \mysize by -20pt 
   \def\epsfsize##1##2{\ifdim##1>\mysize\mysize\else##1\fi}
   \hfill \epsffile{\figname.eps} \hfill 
        } }
 \vbox to 7pt {\ }
 }
\def\mydot{\rule{0.25pt}{0.25pt}}
\def\fignum#1{ \mbox{} }
\def\gloscale{1.0}
\def\dispfig#1{
  \def\figname{#1}
  \def\epsfsize##1##2{\gloscale##1}
  \settowidth{\mysize}{ \epsffile{\figname.eps} }
  \parbox{\mysize}{ \epsffile{\figname.eps} }
  }
%
%

%
%
Although being under consideration for long,  the
strong-coupling-perturbation theory for electronic properties of
High-T$_c$ Superconductors and Heavy-Fermion Systems poses major
difficulties, since it apparently cannot be based on a linked-cluster
expansion with small (i.e.\ one- and two-particle) vertices. 
A perturbation series
in e.g.\ the hybridisation $V$ of the periodic Anderson model either
involves (Hubbard-) cumulants 
\cite{hub66,barcha93} or other kinds 
of vertices with an arbitrary number of legs 
\cite{izylet90}\,, or the linked-cluster theorem is lost in an unconventional
diagrammatic expansion which uses small vertices
\cite{grekei81,col84,kur85,gre87}\,. In the following it is shown
that a Feynman-type expansion around the Atomic Limit may
become available if the problem is restricted to the case of zero
temperature. 

We consider the periodic SU($N$) Anderson model with interaction
$U\to\infty$ of localised $f$-electrons in
auxiliary particle (`Slave Boson') representation, 
$\displaystyle H = 
   H_0 + H_V $
with 
\begin{eqnarray}    \label{eq-hamnull}
  H_0 & = &
    \sum_{k,m} \varepsilon_{k} c^\dagger_{k m} c_{k m} + 
    \sum_{\mu,m} \varepsilon^f s^\dagger_{\mu m} s_{\mu m} 
    \;,\; 
    \\  \label{eq-hamvau}
  H_V & = &  
    \frac{1}{\sqrt{N_L}}\sum_{\mu,k,m}
    \left( V_k e^{-i k R_\mu} s^\dagger_{\mu m} b_\mu c_{k m} + h.c. \right)
    \;\,. 
\end{eqnarray}
Bosons $b_\mu$ and Fermions $s_{\mu m}$ with spin $m$ on $N_L$ lattice
sites $R_\mu$ are subject to operator constraints $Q_\mu=1$ 
for conserved `charges' 
$  Q_\mu = 
    b^\dagger_\mu b_\mu + 
    \sum_m s^\dagger_{\mu m} s_{\mu m}$\,.
Consider a grand canonical `$Q$-ensemble' with 
chemical potential $\lambda$ for the total `charge'
$\widehat{Q} = \sum_\mu Q_\mu$\,, that is
$\displaystyle   H \to K = H - \lambda\widehat{Q}$\,.
The constraints are faithfully represented by
\begin{eqnarray}   \label{eq-const1}
  \langle Q_\mu \rangle^\lambda & = & 1  
    \;,\; \mu=1,\ldots,N_L
    \\    \label{eq-const2}
  \Delta Q_\mu^2 = 
    \langle Q_\mu^2 \rangle^\lambda - \left(\langle Q_\mu
      \rangle^\lambda\right)^2 & = & 0  
      \;,\; \mu=1,\ldots,N_L
\end{eqnarray}
with expectation values in the `$Q$-ensemble'\,. 
Due to the lattice symmetry of the Hamiltonian $K$ the density 
$\displaystyle \langle Q_\mu \rangle^\lambda = 
   \langle \widehat{Q} \rangle^\lambda / N_L$
is site-independent, and Eqs.(\ref{eq-const1}) are fulfilled for
a suitably chosen $\lambda=\lambda(T)$\,. Also is the local fluctuation
$\Delta Q_\mu^2$ site-independent, but for finite temperature it is
$>0$\,.  In the limit $T\to 0$\,, however, fluctuations of conserved
quantities $Q_\mu$ vanish, and Eqs.(\ref{eq-const2}) are 
strictly fulfilled if a unique ground state is assumed. The latter has
to be confirmed explicitely by calculation of $\Delta Q_\mu^2$ at
$T\to 0$\,, since in principle it cannot be ruled
out that $K$ takes a degenerate ground state with fixed
$\widehat{Q}=N_L$\,, i.e.\ an  invariant (reducible) subspace of the 
lattice symmetry containing
states which obey an inhomogeneous distribution of $\widehat{Q}$ on the
lattice with 
$\Delta Q_\mu^2 >0$\,, whereas  $\langle Q_\mu \rangle^\lambda =
1$\,. 

\begin{figure}
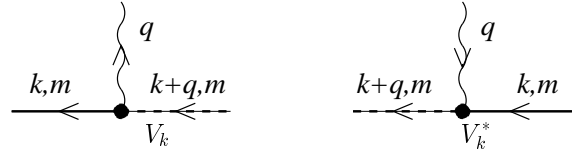

\loadepsfig{figvert}
 %
%
 \caption[\ ]{ 
Vertices for correlated hybridisation events \protect$\sim
V\protect$ in \protect$k\protect$-space. 
         }
 \label{fig-vert}
\end{figure}

The retarded one-particle Green's function $F_{k\,m}(\omega)$ for
$f$-electrons is given by the analytic continuation $i\omega_l \to
\omega + i0_+$ of the fermionic Matsubara Green's function 
\begin{equation}  \label{eq-fgf1}
  F_{k\,m}(i\omega_l) = 
    - \int_0^{\frac{1}{k_BT}}\!\!\!d\tau\,
    \langle {\cal T}\{ \widetilde{ X}^{0 m}(k;\tau)
                       \widetilde{ X}^{m 0}(k; 0)  \} \rangle^\lambda
    \;,\; 
\end{equation}
and the limit $T\to 0$\,. The operator
$\widetilde{ X}^{0 m}(k) = 
   \frac{1}{\sqrt{N_L}}\sum_\mu e^{-i k R_\mu}
   (b^\dagger_\mu s_{\mu m}) $
represents physical $f$-excitations. Similarly the conduction-electron
propagator at $T\to 0$ is accessed through
$G^c_{k\,m}(\tau) = 
  - \langle {\cal T}\{ c_{km}(\tau) c^\dagger_{km}(0) \}
    \rangle^\lambda$\,.
These propagators obey a skeleton Feynman-diagram expansion in
$k$-space with respect to $V_k$\,, using the vertices shown in Fig.\
\ref{fig-vert} and renormalised Matsubara-Green's functions for
auxiliary Fermions (dashed line) 
$ G^{s}_{k\,m}(i\omega_l) = 
   \left( i\omega_l - \varepsilon^f +
   \lambda - \Sigma^s_{k\,m}(i\omega_l)\right)^{-1}$\,,
auxiliary Bosons (wavy line)
$ D_{k}(i\nu_l) = 
   \left( i\nu_l + \lambda - \Pi_k(i\nu_l) \right)^{-1}$\,,
and the conduction electrons (continuous line)
$ G^{c}_{k\,m}(i\omega_l) = 
   \left(i\omega_l - \varepsilon_k - \Sigma^c_{k\,m}(i\omega_l) 
   \right)^{-1}$\,,
with standard diagram rules. 
%

%
%

%

%

%
\begin{figure}
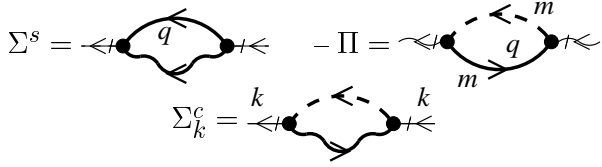
 
\loadepsfig{figself}
 %
%
%
 \caption[\ ]{ 
Self-consistent \protect$(1/N)^1\protect$
approximation for the Anderson lattice. Internal loop spin
\protect$m\protect$ and 
wave vector \protect$q\protect$ are summed. Auxiliary particle
(dashed and wavy) lines are always wave-vector independent (see text).
         }
 \label{fig-self}
\end{figure}
A diagrammatic large-$N$ expansion scheme \cite{bic87} is now easily
applied to the lattice model: In Fig.\
\ref{fig-self} a $\Phi$-derivable self-consistent 
approximation is shown, which contains all diagrams to the 
$f$-Green's function Eq.(\ref{eq-fgf1}) up to order $(1/N)^1$\,.
(Note that $F_{k\,m}$ is written as a geometric series in
$\Sigma^c_{k\,m}$\,.)  
We do not allow for a spontaneous breakdown of gauge symmetry here
(no `spurious Bose condensation'), since the constraint
Eqs.(\ref{eq-const2}) would be massively violated in this case
\cite{millee87}\,. Accordingly Mean-Field (MF) Theory
\cite{newrea87} is
ruled out, Fig.\ \ref{fig-self} represents the simplest
gauge-invariant approximation. Also due to the local $Q$-conservation 
$\displaystyle [ K, Q_\mu ] = 0$
a `$Q$-excitation' on the lattice site $\mu$ in e.g.\ the Boson
propagator 
$\displaystyle D_{\mu'\mu}(\tau) = 
   - \langle {\cal T}\{ b_{\mu'}(\tau) b^\dagger_\mu(0) \}
     \rangle^\lambda
     \sim \delta_{\mu'\,\mu}$
has to be removed at the same site. Therefore the auxiliary
particle's Green's functions and self energies  may not develop a
dispersion, in striking contrast to 
MF Theory. On the other hand, physical quantities like
 $\Sigma^c_k$ are wave-vector
dependent by the matrix element $V_k$ or through vertex
corrections in higher orders $1/N$\,. 

In the $(1/N)^1$ approximation the conduction electron enters only
locally, i.e.~
%
%
$|V_q|^2 G^c_q \to |V|^2 G^c = \frac{1}{N_L}\sum_q |V_q|^2 G^c_q$
in $\Sigma^s$ and $\Pi$\,. For a structure-less hybridisation $V_k=V$
also $\Sigma^c_k$ becomes $k$-independent, leading to a purely local
self-consistency problem given by the self-energies of Fig.\
\ref{fig-self}\,, Dyson's equations for $G^s$ and $D$ with index $k$
omitted, and 
\begin{equation}  \label{eq-gcloc}
  G^c(i\omega_l) = 
    {\textstyle \frac{1}{N_L}}\sum_k\left[ i\omega_l - \varepsilon_k - 
                              \Sigma^c(i\omega_l) \right]^{-1}
    \;\,.
\end{equation}
In order $(1/N)^1$ the lattice model appears as an {\em effective}
single impurity 
Anderson model (SIAM): The reasoning given so far may be repeated for
the SIAM, leading to the saddle-point results of Ref.\ 
\cite{kroetal92} at $T\to 0$\,. The impurity's
self-consistency equations are those of the lattice, solely
Eq.(\ref{eq-gcloc}) is replaced by 
\begin{equation}  \label{eq-gcimp}
  G^c_{\mbox{\tiny SIAM}}(i\omega_l) = 
    \left[ \left( G^{c(0)}_{\mbox{\tiny SIAM}}(i\omega_l)
           \right)^{-1}
           - \Sigma^c_{\mbox{\tiny SIAM}}(i\omega_l)
    \right]^{-1} 
    \;\,.
\end{equation}
The {\em effective} SIAM\,, which reproduces the PAM to
order $(1/N)^1$\,, is defined through the bare conduction-electron
propagator 
$G^{c(0)}_{\mbox{\tiny SIAM}}$
that fulfills Eq.(\ref{eq-gcimp}) with
$\Sigma^c_{\mbox{\tiny SIAM}} = \Sigma^c$\,,
$G^c_{\mbox{\tiny SIAM}} = G^c$\,.

This result also follows from the
limit $d\to \infty$\,: The spatial dimension $d$ 
affects merely the conduction-electron line $G^c_k$\,, 
but only its local part $G^c$ enters the
self energies to order $(1/N)^1$\,. It is shown via e.g.\ a cumulant
representation 
\cite{kur85,hueqin94,jbepl} that in general for $d\to\infty$ 
$\Sigma^c_k$  becomes an {\em effective} local
$\Sigma^c_{\mbox{\tiny SIAM}}$\,, subject to the requirement 
$G^c_{\mbox{\tiny SIAM}} = G^c$\,.
For finite $d$ this defines the Local Approximation.
Accordingly at $V_k=V$ the self-consistent $(1/N)^1$ theory is
equivalent to a Local Approximation. 

The self-consistency equations for the lattice model were iterated
numerically at finite $T$ below the Kondo energy $T_K$ while adjusting
$\lambda$ to fulfill Eq.(\ref{eq-const1})\,. A pseudo gap tends to
appear in the Abrikosov-Suhl resonance in the local $f$-spectrum
$\frac{1}{N_L}\sum_k(-\frac{1}{\pi})\mbox{Im}\,F_k(\omega)$\,.
However, to be able to extrapolate to $T\to 0$ much lower temperatures
have to be reached \cite{sces96-note1}\,. Note that 
$G^{c(0)}_{SIAM}$
needs not be determined here, in contrast to known strong-coupling
approaches applied in the Local Approximation
\cite{kimkurkas90,bri96pre}\,. 

In conclusion it has been demonstrated that restriction to the
case of zero temperature opens the way to a Feynman-type
strong-coupling-perturbation approach to physical Green's functions of
the periodic $U\to\infty$ Anderson model. Since the excluded-volume
problem does not show up here, 
$\Phi$-derivable self-consistent approximations are formulated
conveniently in terms of skeleton diagrams in $k$-space.
To order $(1/N)^1$ the skeleton expansion is proven equivalent to
the Local Approximation. In higher orders non-local effects will appear
through vertex corrections.
\vspace*{5mm}

The author wishes to thank N.~Grewe, J.~Keller, Th.~Pruschke and
J.~Kroha for intensive and fruitful discussions. He also gratefully
acknowledges the hospitality of the ``Graduiertenkolleg:
Komplexit{\"a}t in Festk{\"o}rpern'' at Universit{\"a}t Regensburg.
%
%
%
%

%

%

%
\vspace*{1cm}

 Jan Brinckmann \\
 Department of Physics, Room 12-105 \\
 Massachusetts Institute of Technology \\
 77 Massachusetts Av. \\
 Cambridge, MA 02139 \\
 USA \\
\ \\
 {\bf Fax.:} ++1-617-253-2562 \\
 {\bf Phone:} ++1-617-253-2895 \\
 {\bf e-mail:} {\tt janbri@mit.edu}

\begin{thebibliography}{10}
\bibitem{sces96-note1}
The self-consistency equations may be written without any Bose function for
  finite $T$ , thus obeying a well defined limit $T\to 0$\,.
\end{thebibliography}
\end{document}